\documentclass[aps,prb,reprint,groupedaddress,superscriptaddress,twocolumn]{revtex4-1}
\usepackage{graphicx,color,bm,amsmath,amssymb,natbib,hyperref}

\begin{document}

\title{Topological energy barrier for skyrmion lattice formation in MnSi} 

\author{A.~W.~D.~Leishman}
\affiliation{Department of Physics, University of Notre Dame, Notre Dame, IN 46556, USA}

\author{R.~M.~Menezes}
\affiliation{Department of Physics, University of Antwerp, Groenenborgerlaan 171, B-2020 Antwerp, Belgium}
\affiliation{Departamento de F\'{\i}sica, Universidade Federal de Pernambuco, Cidade Universit\'{a}ria, 50670-901, Recife-PE, Brazil}

\author{G.~Longbons}
\affiliation{Department of Physics, University of Notre Dame, Notre Dame, IN 46556, USA}

\author{E.~D.~Bauer}
\affiliation{Los Alamos National Laboratory, Los Alamos, NM 87545, USA}

\author{M.~Janoschek}
\affiliation{Los Alamos National Laboratory, Los Alamos, NM 87545, USA}
\affiliation{Laboratory for Neutron and Muon Instrumentation, Paul Scherrer Institute, CH-5232 Villigen, Switzerland}

\author{D.~Honecker}
\affiliation{Institut Laue-Langevin, 71 avenue des Martyrs, CS 20156, F-38042 Grenoble cedex 9, France}

\author{L.~DeBeer-Schmitt}
\affiliation{Large Scale Structures Group, Neutron Sciences Directorate, Oak Ridge National Laboratory, Oak Ridge, TN 37831, USA}

\author{J.~S.~White}
\affiliation{Laboratory for Neutron Scattering and Imaging, Paul Scherrer Institut, CH-5232 Villigen, Switzerland}

\author{A.~Sokolova}
\affiliation{Australian Centre for Neutron Scattering, Australian Nuclear Science and Technology Organization, NSW 2234, Australia}

\author{M.~V.~Milo\u{s}evi\'{c}}
\affiliation{Department of Physics, University of Antwerp, Groenenborgerlaan 171, B-2020 Antwerp, Belgium}
\affiliation{NANOlab Center of Excellence, University of Antwerp, B-2020 Antwerp, Belgium}

\author{M.~R.~Eskildsen}
\email[email: ]{eskildsen@nd.edu}
\affiliation{Department of Physics, University of Notre Dame, Notre Dame, IN 46556, USA}

\date{\today}

\begin{abstract}
\noindent

We report the direct measurement of the topological skyrmion energy barrier through a hysteresis of the skyrmion lattice in the chiral magnet MnSi.
Measurements were made using small-angle neutron scattering with a custom-built resistive coil to allow for high-precision minor hysteresis loops.
The experimental data were analyzed using an adapted Preisach model to quantify the energy barrier for skyrmion formation and corroborated by the minimum-energy path analysis based on atomistic spin simulations.
We reveal that the skyrmion lattice in MnSi forms from the conical phase progressively in small domains, each of which consisting of hundreds of skyrmions, and with  an  activation  barrier of several eV.
\end{abstract}

\maketitle

\section{Introduction}

Magnetic skyrmions are topological spin structures that have garnered much attention as they show promise as bits in next generation memory devices~\cite{Fert:2017bt}.
A key ingredient for their stabilization is broken inversion symmetry, 
either in the underlying crystal lattice of bulk magnetic materials or in the interfaces of thin film heterostructures.
This broken symmetry, combined with a strong spin-orbit coupling, produces an antisymmetric exchange interaction known as the Dzyaloshinskii-Moriya interaction (DMI)~\cite{Dzyaloshinsky:1958br, Moriya:1960go}.
More recently there have also been reports of skyrmions stabilized by magnetic frustration~\cite{Kurumaji:2019wn,Hirschberger:2019vn}.

In chiral helimagnets such as MnSi and FeGe, the DMI competes with the exchange interaction to produce three distinct magnetic phases below the Curie temperature, including the skyrmion lattice (SkL) hosting A phase~\cite{Muhlbauer:2009bc,Bauer:2012cw, Moskvin:2013kf}.
The A phase is bounded by first order transitions to the paramagnetic phase on the high temperature side and to the conical phase, where the spins precess around a helix with its axis parallel to the applied field, in all other directions of the field-temperature phase diagram~\cite{Bauer:2012cw}.

Due to the skyrmions' inherent topological structure, there is an energy barrier for both the creation and destruction of the SkL from any non-topological phase (e.g. the conical, helical, or field-polarized ferromagnetic phases).
As a result, both the conical and the SkL phases are bistable as local minima in the free energy over a finite region of parameter space,  giving rise to  phenomena such as quench metastability and field history dependence~\cite{Karube:2016bsa,Makino:2017hh,Bannenberg:2017ws, Oike:2016da, Nakajima:2017uc, Karube:2017wf}.
Unique skyrmionic spin structures have even been predicted to be bistable with each other in certain thin film systems~\cite{Buttner:2018wl}.
The metastability gives rise to activated behavior reported for Fe$_{1-x}$Co$_x$Si~\cite{Wild:2017hj} and Zn-doped Cu$_2$OSeO$_3$~\cite{Birch:2019us}, and the activation barrier for the destruction of a metastable SkL in the latter compound was previously determined from time-dependent measurements~\cite{Wilson:2019ud}.
Similarly, the activation barrier for single skyrmions in magnetic thin films have been predicted from theoretical calculations~\cite{Rohart:2016fw,Lobanov:2016ht,stosic2017paths,Heil:2019kt}.
It is the inherent stability provided by the topological energy barrier that makes skyrmions promising candidates for memory applications, and understanding the nature of this barrier is the key to the development of new skyrmion-based devices.
In spite of this need, a complete description of the nucleation mechanism of the SkL in chiral magnets has not yet been fully established.
 
Here we report direct evidence of the skyrmions' topological energy barrier through a measurement of hysteresis in the SkL-conical phase transition in MnSi, using small-angle neutron scattering (SANS)~\cite{Muhlbauer:2019jt}.
Importantly, these measurements were performed on the equilibrium SkL phase rather than metastable configurations as discussed above.
The existence of hysteresis is direct evidence of the bistability of the SkL and conical phases. We further employ a minimum-energy path analysis, based on an atomistic spin model, to both understand and quantify the nature of this barrier and the microscopic dynamics of the phase transition itself.
The combined data show unambiguously that it is energetically favorable for the SkL phase to form progressively, in domains consisting of hundreds of skyrmions.

\section{Experimental Details}
Initial, exploratory small-angle neutron scattering measurements were performed on the CG-2 General Purpose SANS instrument~\cite{Heller:2018} at the High Flux Isotope Reactor at Oak Ridge National Laboratory, and the D33 instrument at Institut Laue-Langevin~\cite{5-42-477}.
Systematic SANS measurements of the SkL hysteresis were carried out at the SANS-I instrument at the Paul Scherrer Institute (PSI) (neutron wavelength and bandwidth: $\lambda = 0.6$~nm, $\Delta \lambda/\lambda = 10\%$) and the Bilby instrument~\cite{Sokolova:2019jd} at the Australian Nuclear Science and Technology Organization (ANSTO) ($\lambda = 0.5$~nm, $\Delta \lambda/\lambda = 10\%$).
From the beam collimation and the neutron wavelength and bandwidth, one can estimate the experimental resolution~\cite{Pedersen:1990aa}:
\begin{eqnarray}
  \sigma_R^2 & = & 4\pi^2 (\delta \theta/\lambda)^2 + q^2 (\Delta \lambda/\lambda)^2,   \label{RadialRes}\\
  \sigma_L^2 & = & (q \lambda/2\pi)^2 \; \sigma_R^2.   \label{LongRes}
\end{eqnarray}
Here $\sigma_R^2$ and $\sigma_L^2$ are radial and longitudinal resolution respectively, $\delta \theta$ is the standard deviation of the beam divergence, and $q$ is the magnitude of the scattering vector.

The $3.2 \times 2.0 \times 1.3$~mm$^3$ MnSi single crystal used for the SANS measurements was cut from a larger crystal grown by the Bridgman-Stockbarger method followed by a one week annealing at 900~$^\circ$C in vacuum.
The parent crystal has previously been well characterized  confirming its high quality.
Specifically, different pieces of the same crystal were investigated by AC magnetic susceptibility and electrical resistivity measurements~\cite{Fobes:2017ug,Luo:PC}.
This confirmed that the phase diagram agrees well with those reported in literature~\cite{Muhlbauer:2009bc} ($T_{c} = 29$ K), and yielded a residual resistivity ratio (RRR) of 87 (defined as the ratio of the electrical resistivity at 300~K and 2~K).
This is comparable to samples used in previous neutron scattering studies on the SkL in MnSi.
Further pieces were also characterized by resonant ultra sound measurements and energy dispersive X-ray spectroscopy, with the latter confirming the correct stoichiometry~\cite{Luo:PC,Luo:2018tg}.
Finally, an earlier SANS study of influence of uniaxial strain on the SkL has been carried out on parts of the same crystal~\cite{Fobes:2017ug}.
For the SANS experiments, the MnSi crystal was aligned with the $[1 1 0]$ direction parallel to both the applied field and the incident neutron beam, such that only one SkL orientation was energetically favorable, with SkL vector parallel to the crystallographic $[1 \bar{1} 0]$ direction.

At the beginning of each SANS experiment, temperature sweeps (26--32~K) and field sweeps (130--240~mT) were performed to locate the A phase boundaries.
The main SANS results consist of hysteresis loops, obtained by sweeping the field between the SkL and conical phases at constant temperatures. For these loops, temperatures were selected which correspond to the maximal observed scattered intensity of the SkL (28.1 K), and to a 50\% reduction of this intensity on the warmer (28.4 K) and cooler (27.8 K) sides of the A Phase.
For the major loops, the field was swept between 130 and 240~mT using the superconducting cryomagnet. This traverses the entire A phase, with both field endpoints well within the conical phase.
For the minor hysteresis loops a resistive coil was used to supplement to the superconducting magnet, and achieve a higher precision of the magnetic field.
A Cernox sensor and a nichrome heater were mounted in direct contact with the sample disk, allowing an independent temperature control of the sample to within $\pm 10$~mK throughout the minor loops.
Prior to the minor hysteresis loops the sample was heated to a temperature above the A phase, and then field-cooled to the center of the upper phase transition in a constant field of 205~mT.

\section{Experimental Results}

\begin{figure}
    \includegraphics[width=0.8\columnwidth]{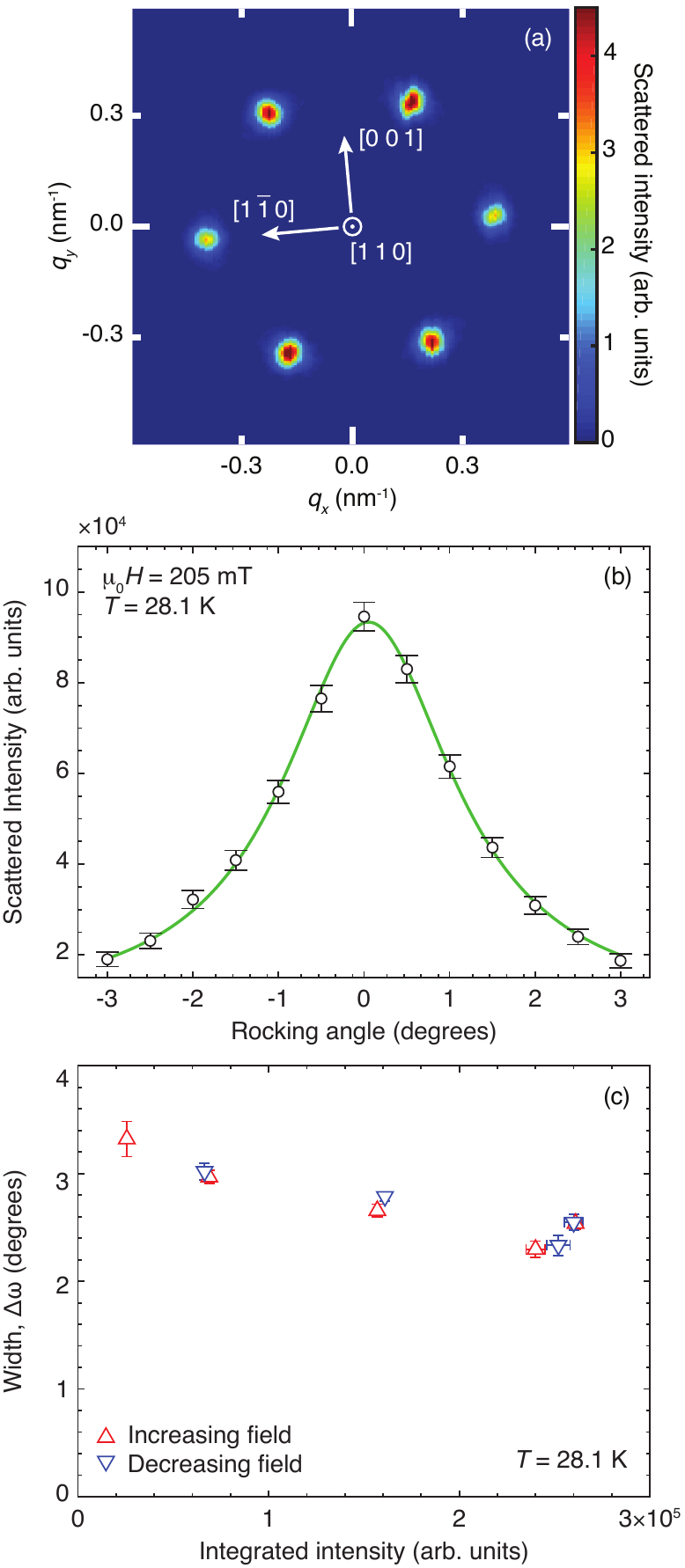}
    \caption{\label{DiffPat}
        (a) Diffraction pattern of the SkL of MnSi at $H = 195$ mT.
        This is a sum of measurements at different rocking angles, with peaks on the horizontal axis appearing fainter as they were, on average, further from the Bragg condition.
        Background scattering near the detector center ($q = 0$) is masked off.
        (b) Rocking curve at $H = 205$~mT, midway along the upper SkL-conical phase transition.
        The curve is fit to a Lorentzian distribution with a width $\Delta \omega = 2.44^{\circ} \pm 0.04^{\circ}$ FWHM.
        (c) Widths, obtained from Lorentzian fits to the rocking curves, along the upper SkL-conical phase transition for both increasing and decreasing field sweeps.}
\end{figure}

A typical SkL diffraction pattern is given in Fig.~\ref{DiffPat}(a).
This shows the sum of the scattered intensity as the SkL is rotated about the vertical axis to satisfy the Bragg condition for each of the six peaks.
Bragg peaks associated with the conical phase are not visible in this geometry, and therefore do not contribute to the scattering.
Figure~\ref{DiffPat}(b) shows the angular dependence of the intensity of a single peak, as both the sample and applied field are rotated together through the Bragg condition.
The rotation is performed perpendicular to the Ewald sphere, eliminating the need for a Lorentz correction of the angular peak width.
This so-called rocking curve is well fitted by a Lorentzian line shape, indicating that it is dominated by spatial or temporal fluctuations of the SkL rather than experimental resolution~\cite{Muhlbauer:2019jt}.
We believe it unlikely that these fluctuations are temporal like those associated with critical fluctuations observed above $T_c$~\cite{Grigoriev_2005,Janoschek_2013,Kindervater:2019kc}, but rather are a result of a finite skyrmion correlation length along the field direction due to crystal mosaicity as reported in other studies of MnSi\cite{Muhlbauer:2009bc}.

Full width half maximum (FWHM) rocking curve widths, $\Delta \omega$, obtained from the Lorentzian fits, are shown in Fig.~\ref{DiffPat}(c).
Here the horizontal axis is the integrated intensity, where the maximal value corresponds to being fully in the SkL phase and zero corresponds to being fully in the conical phase.
The lowest intensity where complete rocking curve measurements are feasible is roughly one tenth of the maximal intensity. 
From the widths one can estimate the longitudinal correlation length $\zeta_L = 2(q_{\text{SkL}} \, \Delta \omega)^{-1}$, where $q_{\text{SkL}} = (0.388 \pm 0.002)$~nm$^{-1}$ is the magnitude of the SkL scattering vector.
As the rocking curve widths greatly exceed $\sigma_R/q_{\text{SkL}} = 0.3^{\circ}$~FWHM obtained from Eq.~(\ref{LongRes}), corrections to $\Delta \omega$ due to the experimental resolution are negligible.
The measured widths yield values of $\zeta_L$ ranging from 130~nm to 90~nm, indicating a reduction of the average SkL domain length along the field direction by the introduction of conical phase regions within the sample.
Similarly, the lateral correlation length $\zeta_R$ can be estimated from FWHM of the Bragg peak in the radial direction within the detector plane, $\Delta q_R$.
Fits of the radial intensity yields $\Delta q_R \sim 6.3  \times 10^{-2}$~nm$^{-1}$ fully within the SkL phase, increasing to $\sim 6.9 \times 10^{-2}$~nm$^{-1}$ upon entering the conical phase [apart from a re-scaling of the vertical axis, the behavior is near identical to that of the rocking curve widths in Fig.~\ref{DiffPat}(c)].
Correcting for the comparatively poorer resolution within the detector plane yields $\zeta_R = 2 (\Delta q_R^2 - \sigma_R^2)^{-1/2}$, with $\sigma_R = 4.8 \times 10^{-2}$~nm$^{-1}$ obtained from Eq.~(\ref{RadialRes}).
From this, one finds a lateral correlation length decreasing from 50 to 40~nm upon leaving the SkL phase.
Together, these results suggest that the phase transition proceeds locally, with nanoscale regions transitioning independently over a range of applied fields.

The total integrated Bragg peak intensity is proportional to the number of scatterers (skyrmions) in the system~\cite{Muhlbauer:2019jt}, and therefore the fraction of the sample volume within the SkL phase.
Within the detector plane integration is performed by summing counts in the pixels spanning a Bragg peak.
Integration along the third dimension of reciprocal space is obtained from rocking curves.
However, as the applied field $H$ is increased into the conical phase and the scattered intensity from the SkL vanishes, the rocking curve widths only change modestly as seen in Fig.~\ref{DiffPat}(c).
The SkL volume fraction is thus taken to be proportional to the rocking curve peak intensity for studies of hysteresis associated with the SkL-conical phase transition.
While it is possible to make corrections for the systematic variation in the rocking curve width in Fig.~\ref{DiffPat}(c), this is a comparatively small effect and does not influence the analysis of the data in a significant manner as we shall discuss later.

\begin{figure}
    \includegraphics[width=0.8\columnwidth]{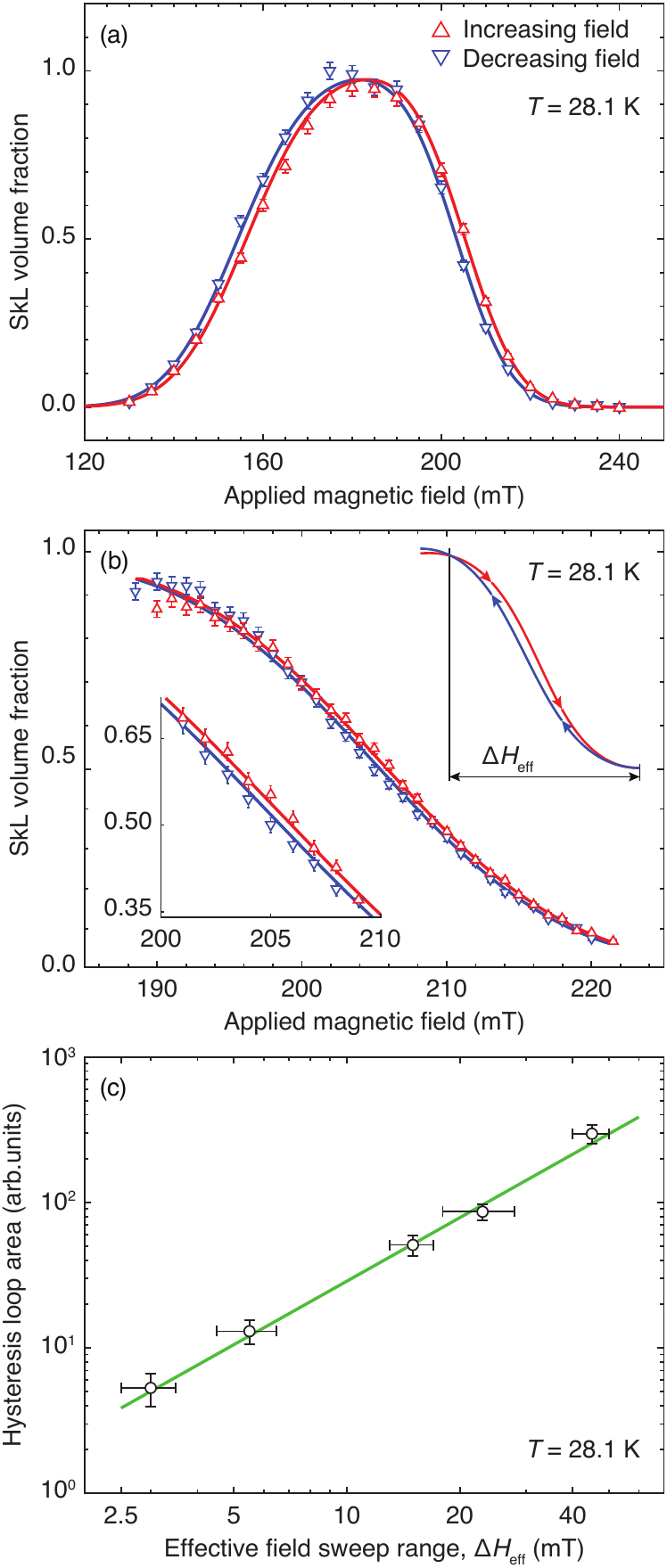}
    \caption{\label{Hyst}
        (a) Major hysteresis loop for $T = 28.1~K$ recorded at PSI.
        (b) Minor hysteresis loop at the same temperature, centered around 205~mT and with a field sweep range of 33~mT. Symbols are the same as in panel (a).
        Bottom left inset: Expanded view of the central part of the loop.
        Top right inset: Schematic showing field sweep direction and effective sweep range $\Delta H_{\text{eff}}$.
        Curves in (a) and (b) are fits to an adapted Preisach model described in the text.
        (c) Area of hysteresis loops as a function of the effective sweep range.}
\end{figure}

Figures~\ref{Hyst}(a) and \ref{Hyst}(b) shows respectively a major and a minor hysteresis loop at $T = 28.1$~K.
In both cases, the intensity was normalized by the maximal observed intensity, which corresponds to the entire sample being in the SkL phase.
In the major hysteresis loop, the field was swept from 130 to 240~mT and back.
Both end points are well inside the conical phase, and this loop covers the entire A phase.
Here, a clear separation of the two sweep directions is observed, with the SkL volume fraction lagging in the direction the field is changing.
Importantly, thermal relaxation times in MnSi are much shorter than the SANS count times at the measurement temperatures~\cite{Oike:2016da}, and do not contribute to the hysteresis.

To confirm hysteretic behavior, a series of minor loops were measured, each of which was centered on the high field phase transition into the conical state.
Prior to each minor loop, the sample was cooled from the paramagnetic state to the measurement temperature in a constant field (205~mT), followed by a reduction of the field to the starting point.
From here, minor hysteresis loops were recorded by raising the field to partially leave the SkL phase and then decreasing it to reenter.
An example of a minor loop is show in Fig. ~\ref{Hyst}(b).
The minor loops show a clear nesting, quantified by the loop area which grows superlinearly as the loops become longer as shown in Fig.~\ref{Hyst}(c).
Here the horizontal axis is the effective field sweep range $\Delta H_{\text{eff}}$, defined as the separation between the two crossing points of the different field sweep directions illustrated in the Fig.~\ref{Hyst}(b) inset.
Values for $\Delta H_{\text{eff}}$ and the loop area were determined by fits to the data described below, and the area was found to grows as a power law $\propto \Delta H_{\text{eff}}^{1.45\pm0.1}$.

\begin{figure}
    \includegraphics[width=\columnwidth]{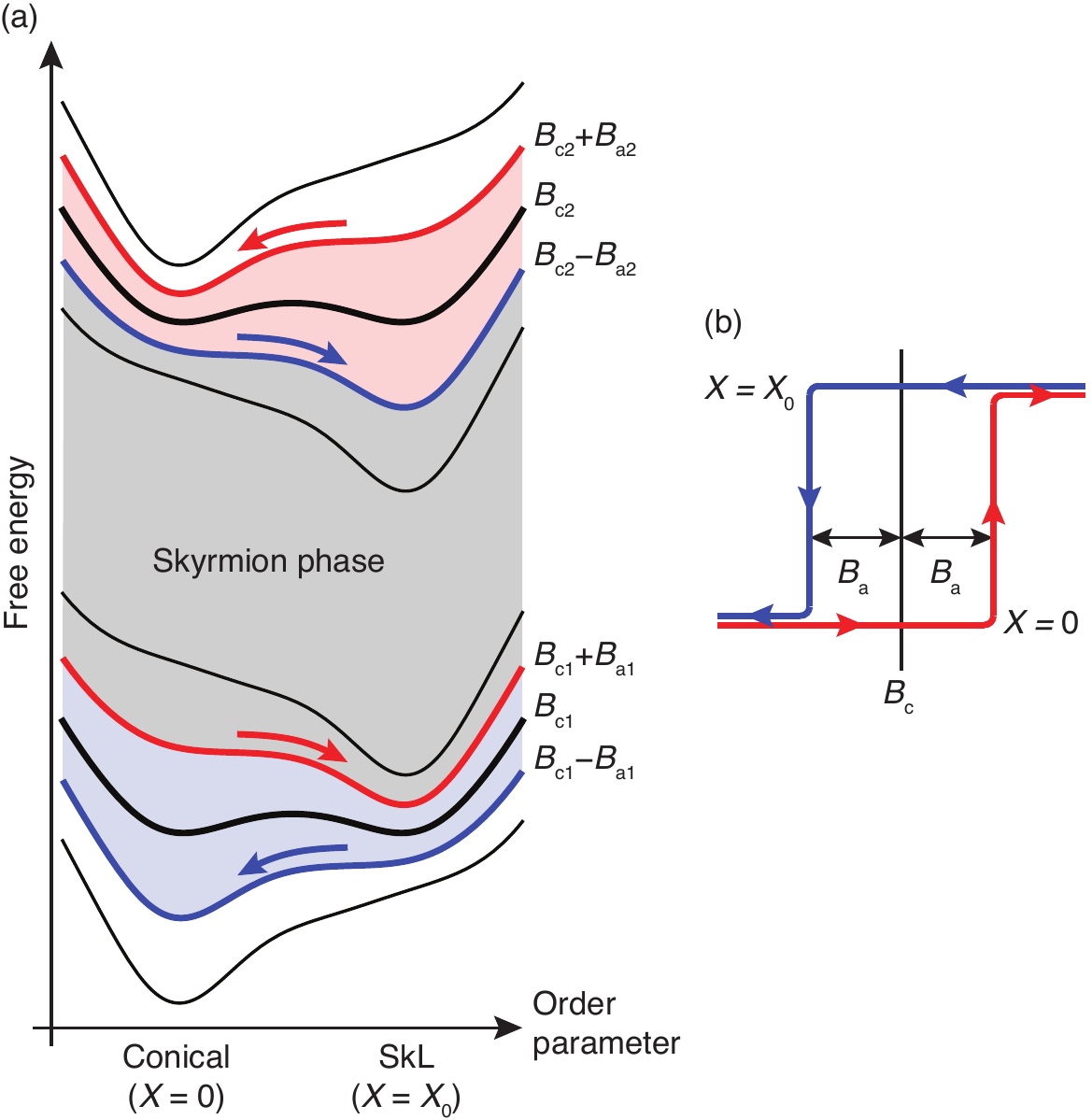}
        \caption{\label{PreisachCurvy}
            Behavior of an individual Preisach unit.
            (a) Free energy for different values of the applied field.
            Black curves correspond to fields where the conical and SkL phases have the same energy.
            Red (blue) curves indicate the location of the phase transition for increasing (decreasing) field.
            (b) Hysteretic response of the order parameter.}
\end{figure}

To quantify the activation barrier for skyrmion formation and destruction, the SANS hysteresis loops are analyzed using an adapted Preisach model. This is suitable for transitions in bistable systems, where two phases coexist as local free energy minima over some range of the external field \cite{Bertotti}.
In the region of bistability, the free energy $F$ is assumed to be linearly proportional to the magnetic field $B$:
\begin{equation}
  \label{FreeEnergy}
  F(B,T,\ldots) = F(B_c,T,\ldots) \mp (X - X_0/2) (B-B_c).
\end{equation}
Here, $X$ is an order parameter with dimensions of a magnetic moment, used to distinguish the conical ($X = 0$) and skyrmion ($X = X_0$) phases.
The sign of the second term in Eq.~(\ref{FreeEnergy}) correspond to respectively the lower (-) and upper (+) transition between the SkL and conical phases.
The Preisach free energy as a function of  applied field is shown in Fig.~\ref{PreisachCurvy}(a).
A similar picture was previously proposed to describe temperature-quenched metastable SkL phases in MnSi~\cite{Oike:2016da}.

\begin{table*}
    \begin{tabular}{lccccccc}
    \hline \hline
    Facility & $T$~(K) & $\overline{H}_{c1}$~(mT) & $\sigma_{c1}$~(mT)  & $\overline{H}_{a1}$~(mT) & $\overline{H}_{c2}$~(mT) & $\sigma_{c2}$~(mT) & $\overline{H}_{a2}$~(mT) \\ \hline \hline
     ANSTO & 27.8 & 188 $\pm$ 8 & 19 $\pm$ 2 & 1.1 $\pm$ 0.3 & 211 $\pm$ 3 & 14 $\pm$ 1 & 1.0 $\pm$ 0.2 \\
    PSI & 28.1 & 155.3 $\pm$ 0.2 & 12.5 $\pm$ 0.2 & 0.94 $\pm$ 0.14 & 204.4 $\pm$ 0.2 & 9.5 $\pm$ 0.2 & 0.96 $\pm$ 0.12\\
    ANSTO & 28.1 & 160.2 $\pm$ 0.5 & 13.7 $\pm$ 0.3 & 1.1 $\pm$ 0.2 & 212.5 $\pm$ 0.4 & 12.0 $\pm$ 0.3 & 0.8 $\pm$ 0.2\\
     ANSTO & 28.4 & 168 $\pm$ 9 & 21 $\pm$ 3 & 1.0 $\pm$ 0.3 & 200 $\pm$ 6 & 19 $\pm$ 2 & 0.7 $\pm$ 0.3  \\ \hline \hline
    \end{tabular}
    \caption{\label{MajorFits}
        Preisach parameters obtained from fits to major hysteresis loops.
        Uncertainties indicate the one sigma confidence interval provided by the fitting algorithm.}
\end{table*}

The low- and high-field transitions are treated independently, with each one governed by a pair of parameters: the critical field ($B_{c1/c2}$) where the two phases have the same free energy, and the height of the activation barrier ($B_{a1/a2}$) that inhibits the transition.
As the external magnetic field is increased from zero and approaches the lower conical-to-SkL phase transition, the conical state free energy increases and the SkL state free energy decreases.
At $B = B_{c1} + B_{a1}$, the conical phase minimum vanishes and the system transitions to the skyrmion phase.
For decreasing fields, the transition occurs at $B = B_{c1} - B_{a1}$.
Similarly, the upper SkL-to-conical transition occurs at $B = B_{c2} \pm B_{a2}$, where the situation is reversed and the conical and SkL free energies respectively decrease and increase with increasing field.
The Preisach model is an inherently zero-temperature model, and a transition between the states only occur when one minima disappears and the system is no longer bistable.
This is appropriate for the SkL as reported activation barriers are much greater than $k_B T$~\cite{Wilson:2019ud} for $T \leq T_{c}$.

Preisach free energy curves produce perfectly rectangular hysteresis loops, centered around $B_c$ and with width $2 B_a$, as shown in Fig.~\ref{PreisachCurvy}(b).
Rounded loops are obtained by considering the sample to be composed of microscopic, independently-acting, ``Preisach units'', each with its own $B_{c1/c2}$ and $B_{a1/a2}$. Since the magnetization is approximately linear across both the upper and lower field phase transitions \cite{Bauer:2012cw}, we express $B_{c1/c2}$ and $B_{a1/a2}$ in terms of the corresponding applied fields $H_{c1/c2}$ and $H_{a1/a2}$.
To model the SANS hysteresis loops, Preisach units are assumed to follow a Gaussian distribution in both critical and activation fields.
These distributions are characterized by their mean values ($\overline{H}_{c1/c2}, \overline{H}_{a1/a2}$) and standard deviations ($\sigma_{c1/c2}, \sigma_{a1/a2}$).

A fit to the PSI major hysteresis loop for $T = 28.1$~K is shown in Fig.~\ref{Hyst}(a), and the resulting parameter values are summarized in Table~\ref{MajorFits}.
Values of $\sigma_{a1/a2}$ converge to zero during the fit, and this parameter was therefore eliminated.
Differences between the fit and the data near the maximum SkL volume fraction are due to the Gaussian Preisach distribution used.
A skewed distribution, introducing additional degrees of freedom, could improve the overall fit.
However, the values of $\overline{H}_{a}$, which is the principal variable of interest, would most likely remain unchanged as they depend on the width of the hysteresis (separation of up- and down-sweeps) at half SkL volume fraction, where the current fits are very good.
Finally, rescaling the data to account for the changing rocking curve width previously discussed only effects the Preisach fits minimally.
Specifically, $\overline{H}_{c1}$/$\overline{H}_{c2}$ are shifted by $\sim 2$\% in opposite directions to increase the width of the SkL phase, $\sigma_{c1}$/$\sigma_{c2}$ are both reduced by $\sim 5$\%, and $\overline{H}_{a1}$/$\overline{H}_{a2}$ remain within uncertainty of the values in Table~\ref{MajorFits}.

Also included in Table~\ref{MajorFits} are results of fits to the major loops recorded at ANSTO at three different temperatures.
The difference in the fitted values of $\overline{H}_{c1/c2}$ at 28.1~K may be attributed to variations in the remnant field of the cryomagnets used, supported by the similar separation between the upper and lower transitions for the PSI and ANSTO results.
The larger uncertainty on $\overline{H}_{c1}$/$\overline{H}_{c2}$ and greater values of $\sigma_{c1}$/$\sigma_{c2}$ at 27.8~K and 28.4~K are due to the weaker scattering at these temperatures.
Importantly, the least affected parameters are the two activation fields, which remains consistent and with modest uncertainties across all the measurements.

As the two transitions are treated independently some Preisach units could, in principle, return to the conical phase before others have entered the SkL phase.
At 28.1~K, where the separation of the transition fields is much greater than $\sigma_{c1/c2}$, this rarely occurs. However, at 27.8 and 28.4~K the transitions overlap significantly, preventing the intensity from reaching the maximum at 28.1~K, which is reflected in the increased values of $\sigma_{c1}$/$\sigma_{c2}$.
More importantly, the good agreement between $\overline{H}_{a1}$ and $\overline{H}_{a2}$ supports a topological origin for the activation barrier which should be similar for both phase transitions.
Further support for this conclusion comes from the comparable values of the activation fields at different temperatures.
This indicates that the finite temperature range of the A phase is not due to a significant reduction of the activation barrier, but rather a convergence of the two critical fields as the energy separation between the conical and SkL phases is reduced.

While applying the Preisach model does not require prior knowledge about the nature of individual units, it is nonetheless relevant to consider their nature.
In the original application to ferromagnetic hysteresis, magnetic domains behave sufficiently independent to be treated as Preisach units. 
By analogy, we anticipate that in the present case they correspond to microscopic SkL domains, within which the cascade of individual skyrmion formation occurs much  faster than the measurement time.
In this way, each domain experiences the phase transition quasi-instantaneously and independent of other domains.
This is consistent with the longitudinal and lateral correlation lengths discussed previously, providing a characteristic length scale for the SkL domains of the order 100 and 50~nm, respectively.
In such a scenario, variations of the local magnetic field due to crystal inhomogeneities and demagnetization effects give rise to a range of different transition fields and therefore a non-zero $\sigma_c$.

It is likely that both the distribution of SkL domains throughout the sample as well as their sizes depend on the field and temperature history, which may affect the activation barriers observed in the SANS experiments.
To explore this possibility Preisach model fits were performed on the minor hysteresis loops, where the initial configuration was obtained by a field cooling to the midpoint of the SkL-conical transition.
In contrast, the major loop has a starting point entirely within the conical phase.
The results of the minor loop fits are summarized in Table~\ref{MinorFits}.
While the values of $\overline{H}_{c2}$ agree with those obtained from the major loop, $\overline{H}_{a2}$ is reduced significantly, confirming that the barrier to create or destroy SkL domains depends on the field history.
We return to this point later.
\begin{table}
    \begin{tabular}{cccc}
    \hline \hline
    $\Delta H_{\text{eff}}$~(mT) & $\overline{H}_{c2}$~(mT) & $\sigma_{c2}$~(mT) & $\overline{H}_{a2}$~(mT) \\ \hline \hline
    5.5 $\pm$ 1.0 & 203.6 $\pm$ 0.2 & 11.5 $\pm$ 0.2 & 0.18 $\pm$ 0.05 \\ \hline
    15 $\pm$ 2 & 204.8 $\pm$ 0.1 & 10.5 $\pm$ 0.1 & 0.16 $\pm$ 0.05 \\ \hline
    23 $\pm$ 5 & 205.1 $\pm$ 0.2 & 10.5 $\pm$ 0.1 & 0.25 $\pm$ 0.04 \\ \hline \hline
    \end{tabular}
    \caption{\label{MinorFits}
        Preisach parameters obtained from minor hysteresis loops at $T = 28.1$~K (PSI).
        }
\end{table}

\section{Theoretical modeling}
\label{TheoryFrame}

To complement the SANS data, atomistic spin dynamics simulations were performed to investigate the transition between the conical and SkL states using a homemade simulation code~\cite{stosic2017paths} as well as the \textit{Spirit} package~\cite{muller2019spirit}.
The extended Heisenberg Hamiltonian that describes the system of classical spins can be written as
\begin{equation}
\begin{aligned}
  \mathcal{H}
    = & -J\sum_{\langle i,j \rangle} \textbf{n}_i \cdot \textbf{n}_j - \sum_{\langle i,j \rangle} \textbf{D}_{ij} \cdot (\textbf{n}_i \times \textbf{n}_j)\\
      & -\sum_i \mu\textbf{B} \cdot \textbf{n}_i,
    \label{HeisenbergHam}
\end{aligned}
\end{equation}
where $\bm{\mu}_i$ is the magnetic moment of the $i^{\text{th}}$ atomic site with $|\bm{\mu}_i| = \mu$, and $\textbf{n}_i = \bm{\mu}_i/\mu$ is the $i^{\text{th}}$ spin orientation.
Here $J$ represents the first-neighbours exchange stiffness, $\textbf{D}_{ij}$ is the DMI vector, $\textbf{B}$ is the perpendicular external magnetic field, and $\langle i,j \rangle$ denotes pairs of nearest-neighbour spins $i$ and $j$.
For the simulations we adopt parameters $J = 1$~meV and $D = 0.18J$, which are reasonable values for MnSi~\cite{iwasaki2013universal,iwasaki2013current}.
Although intrinsic exchange and cubic anisotropies~\cite{Bak_1980} may define a preferential direction for the spin rotation in MnSi at zero field, such high-order contributions are much weaker than the energy terms in Eq.~(\ref{HeisenbergHam}) and are therefore neglected in the calculations.
Similarly, the small contribution from a dipolar interaction is also not included~\cite{Maleyev_2006,Grigoriev_2006}.
The dynamics of the spin system is governed by the Landau-Lifshitz-Gilbert equation
\begin{equation}
  \frac{\partial \textbf{n}_i}{\partial t}
    = -\frac{\gamma}{(1+\alpha^2)\mu_i} \left[\textbf{n}_i \times \textbf{B}_i^\text{eff} + \alpha\textbf{n}_i \times (\textbf{n}_i \times \textbf{B}_i^\text{eff}) \right],
\end{equation}
where $\gamma$ is the electron gyromagnetic ratio, $\alpha$ is the damping parameter and $\textbf{B}_i^\text{eff} = -\partial \mathcal{H}/\partial\textbf{n}_i$ is the effective field. 

\begin{figure}
    \includegraphics[width=5cm]{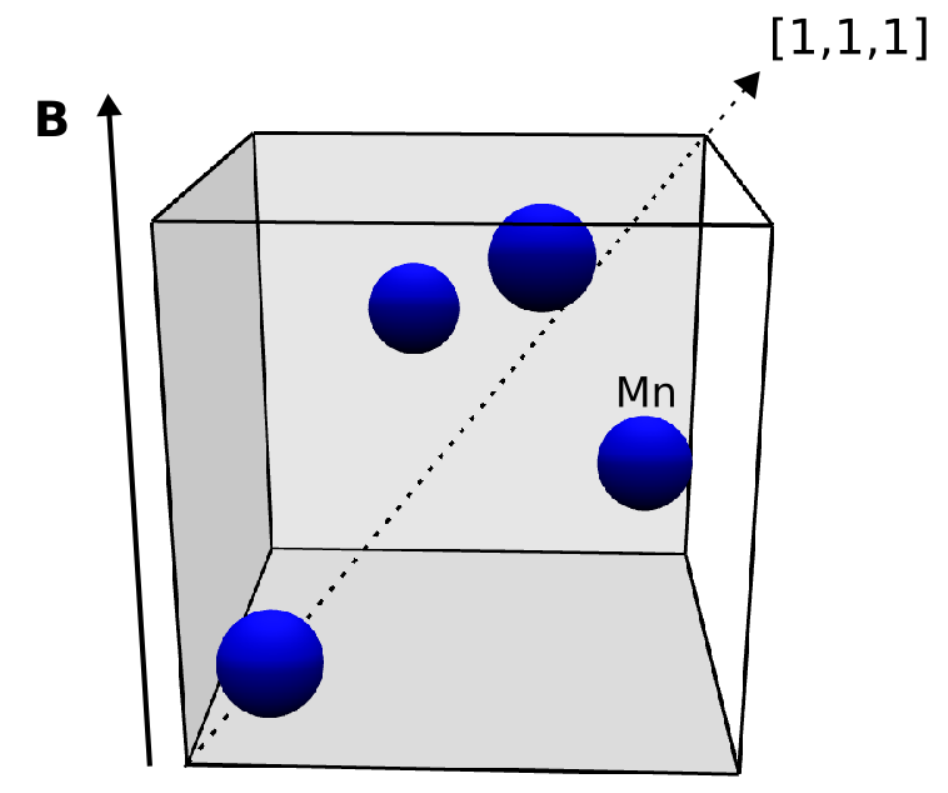}
    \caption{\label{B20struct}
        Unit cell of the B20-structure of MnSi showing only the location of the Manganese atoms.
        The magnetic field $\textbf{B}$ is applied along the $[0 0 1]$ direction.}
\end{figure}

\begin{figure*}
    \includegraphics[width=15cm]{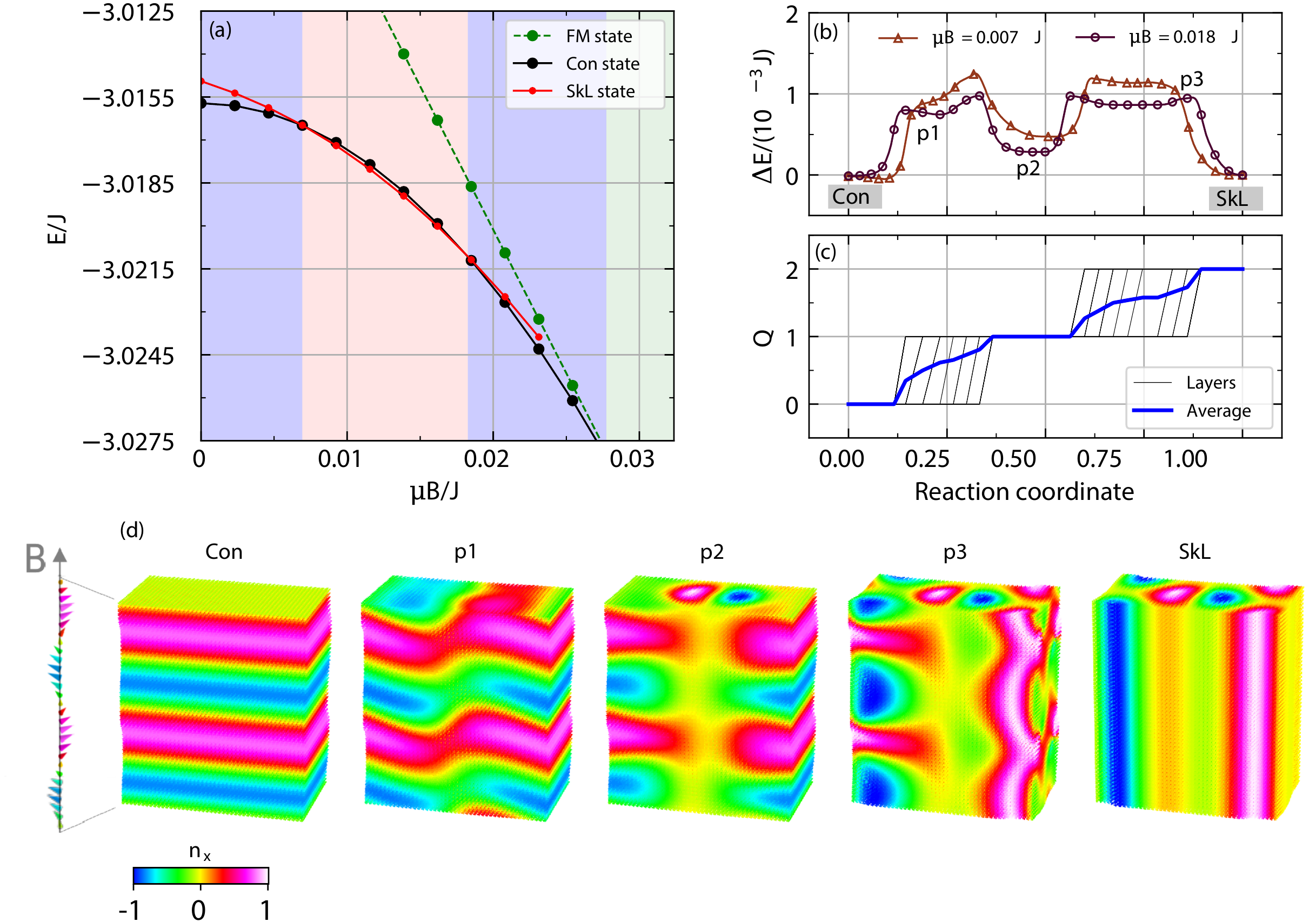}
    \caption{\label{PhaseDiagram}
        (a) Energy per spin vs applied field for each state.
        The ground state is indicated by the colored shading with blue for the conical (Con) state, red for the SkL and green for the field-polarized ferromagnetic (FM) state.
        (b) Minimal energy path between conical and SkL states for $\mu B = 0.007J$ and $\mu B = 0.018J$.
        (c) Topological charge as a function of the reaction coordinate for $\mu B=0.018J$.
        (d) Spin configurations in a $N \times \sqrt{3}N \times 2N$ mesh along the formation path for $\mu B = 0.018J$, as indicated in panel (b) (see also animated data in Ref.~\onlinecite{supplemental}).}
\end{figure*}

The MnSi crystal, shown in Fig.~\ref{B20struct}, consists of a B20 structure (space-group P2$_1$3) with four Mn atoms and four Si atoms located at the $4(a)$-type sites of the simple-cubic unit cell with position coordinates ($u, u, u$), ($0.5+u, 0.5-u, -u$), ($-u, 0.5+u, 0.5-u$), and ($0.5-u, -u, 0.5+u$), where $u_{_\text{Mn}}=0.137$ and $u_{_\text{Si}} = 0.845$~\cite{jeong2004implications}.
For the simulations, only Mn magnetic moments are considered. The spin dynamics simulations were performed in a mesh of $N\times\sqrt{3}N\times N$ unit cells with $N = 26$, and the SkL state consists of two skyrmion tubes located at respectively the center and corners.
The choice of $N$ was verified to minimize the SkL energy.
Periodic boundary conditions are considered along the three dimensions.
To obtain the ground state of the spin model, the energy of the considered states are minimized for different values of the applied field $\textbf{B} \parallel [0 0 1]$.
The choice of field direction parallel to one of the unit cell main axes ensures that skyrmions form as uniform tubes within the simulation box.
However, the direction of the applied field is not expected to have much impact on the energetics as long as a high-symmetry direction of the crystal is chosen.
Figure~\ref{PhaseDiagram}(a) shows the energy obtained in the simulations for the field-polarized ferromagnetic, conical and SkL states, from where the ground state is found to be conical for $\mu B < 0.007J$ and $0.018J < \mu B < 0.028J$, SkL for $0.007J < \mu B < 0.018J$, and field-polarized ferromagnetic for $\mu B > 0.028J$.

Next, the transition between conical and SkL states is considered.
At the critical fields $\mu B_{c1} = 0.007J$ and $\mu B_{c2} = 0.018J$, both states have approximately the same energy.
The activation barrier between the two states can be calculated by the geodesic nudged elastic band (GNEB) method~\cite{bessarab2015method,stosic2017paths} and a climbing image method~\cite{henkelman2000improved}, allowing a precise determination of the highest energy saddle point along the minimal energy path connecting the two states.
Here, the reaction coordinate defines the normalized (geodesic) displacement along the formation path.
Figure~\ref{PhaseDiagram}(b) shows the activation barrier calculated between the two states in both critical fields.
From this one finds that it is energetically favorable to break the conical state locally in different stages, nucleating the skyrmions individually instead of the whole lattice at once (see also animated data in Ref.~\onlinecite{supplemental}).
Figure~\ref{PhaseDiagram}(c) shows the topological charge, given by~\cite{Fert:2017bt}
\begin{equation}
  Q = \frac{1}{4\pi}\int\textbf{n}\cdot(\frac{\partial\textbf{n}}{\partial x}\times\frac{\partial\textbf{n}}{\partial y}) \, dx \, dy,    
\end{equation}
calculated along the formation path for each $xy$-layer of the sample for $B = B_{c2}$.
Notice that the tube of the first skyrmion is formed gradually, layer-by-layer, in a conical background and the average topological charge approaches $Q = 1$, giving rise to the first elongated maximum in the minimal energy path.
This is consistent with previous works suggesting that skyrmions are nucleated or annihilated by the formation and subsequent motion of Bloch points (magnetic monopoles)~\cite{Milde:2013aa,Schutte:2014kh,Rybakov:2015dh}.
After that, the second skyrmion is formed in a similar way, after which the average topological charge approaches $Q = 2$ and the transition is complete.
Energetically equivalent paths were obtained for the first skyrmion nucleating either at the center or the corners.

As recognized previously, the transitions between the SkL and conical states are not expected to occur in a spatially homogeneous fashion.
As a result, the average energy per spin necessary to nucleate a single skyrmion depends on the lateral size of the domains.
An estimation of the activation barrier can be obtained by comparing the energy separation $\Delta E_a = |E_\text{SkL} - E_{\text{Con}}|$ of the SkL and conical states near the critical field, due to an activation field $B_a$ equivalent to the one obtained from the SANS experiments.
Adjusting for the difference between the transition fields obtained experimentally and from the simulations one finds $B_a \approx (B_{c2} - B_{c1})/50 \approx 2 \times 10^{-4}J/\mu $, and from there $\Delta E_a \approx 10^{-5}J$.
This value is roughly two orders of magnitude smaller than the activation energy calculated in the GNEB simulation where the SkL was formed in two steps.
Therefore, to nucleate one skyrmion with a 100 times smaller activation field in the simulations we need to consider a phase transition that occurs in 100 times as many steps as previously.
This is exactly equivalent to using a 100 times larger simulation box, as the activation energy is given by the number of skyrmion nucleations per area.
Considering the SkL periodicity of $19$~nm in MnSi~\cite{Muhlbauer:2009bc}, this corresponds to skyrmion domains of order $\sim 0.05$~$\mu$m$^2$.
This is the same order of magnitude as the correlation length determined directly from the SANS rocking curve widths.

As the formation barrier for the individual skyrmions along the reaction coordinate are all roughly the same height (see Fig.~\ref{PhaseDiagram}(b)), once the system has sufficient energy to overcome the initial barrier skyrmions will continue to nucleate until defects or demagnetization makes it energetically unfavorable.
This limits the size of the SkL domains, and we speculate that this mechanism is responsible for the discrete Preisach units observed in the SANS measurements.
In contrast, the change of SkL volume fraction for the minor hysteresis loops is due to the expansion/reduction of already present domains formed during the field cooling.
This results in a smaller activation barrier, which persists since the crystal never reaches a fully saturated conical or SkL phase throughout the minor loop.
Spatially resolved measurements would be required to confirm this picture. 

The topological energy barrier for each skyrmion can be estimated by multiplying $\Delta E_a$ by the number of spins within a SkL unit cell, and increasing the length of the skyrmions in the simulations to the thickness of the single crystal used in the SANS experiments.
Using the above relationship between $B_a$ and $J/\mu$ with $\mu = 0.4 \mu_B$~\cite{Lonzarich:1985bb}, this
yields $\Delta E_a \approx 7$~eV per skyrmion.
By the nature in which it was obtained, the activation energy above should be considered as an estimate rather than an exact value. Taking into account that $\Delta E_a$ scales linearly with the sample thickness, our estimate for MnSi is roughly 3--4 times greater than the $\sim 1.6$~eV reported for zinc-substituted Cu$_2$OSeO$_3$~\cite{Wilson:2019ud}.
This difference may be due to the higher temperature ($\sim 53$~K versus $\sim 28$~K) and lower fields ($\sim 25$~mT versus $\sim 180$ ~mT) at which the A phase exists in Cu$_2$OSeO$_3$.

\section{CONCLUSION}
\label{Sec:Conclusion}
\noindent
In summary, we presented the first direct observation of the hysteresis in the formation and destruction of the skyrmion lattice in MnSi.
The measured hysteresis proves that the skyrmion lattice and the conical phase are bistable over a finite range of parameters, with a finite \textit{topological} activation barrier inhibiting the phase transition in either direction.
This observation validates the topological stability of skyrmions.
Comparing the experimental data to the results of atomistic spin simulations indicates that the skyrmion lattice is formed progressively in smaller domains, containing hundreds of skyrmions, with an activation barrier of several eV/mm for a single skyrmion. 

Our results advance the understanding of the nucleation mechanism of the SkL in chiral magnets, and we expect that our findings will instigate further measurements of topological energy barriers between different (chiral) magnetic states.
Such studies are key to understanding the evolution of magnetic states in bulk and ultrathin materials and will establish definitively the feasibility of high-density devices based on topological spin structures.

\section*{Acknowledgements}
This work was supported by the U.S. Department of Energy, Office of Basic Energy Sciences, under Award No. DE-SC0005051 (A.W.D.L., G.L., M.R.E.),
the Research Foundation - Flanders (FWO-Vlaanderen) (R.M.M., M.V.M.), and
Brazilian Agencies FACEPE, CAPES and CNPq (R.M.M.).
M.J. was supported by the LANL Directed Research and Development (LDRD) program via the Directed Research (DR) project ``A New Approach to Mesoscale Functionality: Emergent Tunable Superlattices (20150082DR).''
E.D.B. was supported by the U.S. Department of Energy, Office of Basic Energy Sciences, Division of Materials Science and Engineering, under project "Quantum Fluctuations in Narrow-Band Systems".
A portion of this research used resources at the High Flux Isotope Reactor, a DOE Office of Science User Facility operated by the Oak Ridge National Laboratory.
Part of this work is based on experiments performed at the Swiss spallation neutron source SINQ, Paul Scherrer Institute, Villigen, Switzerland.
We acknowledge useful conversations with E.~Louden, D.~Green, and A.~Francisco in preparation for these experiments, as well as the assistance of K.~Avers, G.~Taufer, M.~Harrington, M.~Bartkowiak, and C.~Baldwin in completing them.


%

\end{document}